# Proof that the maximally localized Wannier functions are real

**Sangryol Ri, Suil Ri**

Institute of Physics, Academy of Sciences, Unjong District, Pyongyang, DPR Korea

**Abstract**

The maximally localized Wannier functions play a very important role in the study of chemical bonding, ballistic transport and strongly-correlated system, etc. A significant development in this branch was made in 1997 and conjectured that the maximally localized Wannier functions are real. In this paper, we prove the conjecture. A key to this proof is that the real parts of complex Wannier functions equal to the algebraic average of two real Wannier functions and the spread functional of one of the two real Wannier functions is less than one of complex Wannier functions. If one starts with initial real Wannier functions, the gradients of spread functional to get the maximally localized Wannier functions are only calculated in half of Brillouin zone and more localized Wannier functions are obtained.

## I. INTRODUCTION

Replacing the original many-body problem by an independent-particle problem by Kohn and Sham [1], the electronic ground states of the condensed matter have been successfully solved in terms of a set of Bloch states. Density functional theory is widely used for electronic structure calculations [2], but alternative representations such as Wannier representation are available. The Wannier functions (WFs) give a real-space picture of the electronic structure of a system. The representation in terms of localized WFs was introduced by Wannier [3].

The WFs provide an insightful analysis of the nature of chemical bonding[13,14] and has led to important advances in the characterization and understanding of dielectric response and polarization in materials[15-23]. In particular, the knowledge of the WFs is important in theory of ballistic transport, where Green's functions and self-energies can be constructed effectively in a Wannier basis [4, 5, 24-28].

An important progress in this field was the introduction by Marzari and Vanderbilt about maximal localization criterion for identifying a unique set of WFs [6]. They showed that minimization of a localization functional corresponding to the sum of the second moment spread of each Wannier charge density about its own center of charge was both formally attractive and computational tractable. And they found that whenever they arrive at a global minimum, the WFs always turned out to be real, apart from a trivial overall phase by their empirical experience (conjecture) [6]. In the case of a one-dimensional crystal, it has been proven that the exponentially decaying WFs are real and symmetric about the origin [7]. To the best of our knowledge, it has not been reported any proof of this conjecture for a general three-dimensional system [6, 8-12].

In optics, it is well known that the mode-locked laser pulse is generated when the phases of all the modes coincide. The Bloch wave function is expressed in the similar mathematical form as the electromagnetic wave. From this, it is reasonable to expect that the maximally localized Wannier functions ( MLWFs ) would be always real.

In this paper, we prove that MLWFs are real. In section II, we review some physical concepts necessary for the proof. In section III, we briefly discuss on the global minimum and derive the expression for the spread functional with the plane wave. In section IV, to prove that MLWFs are real, we demonstrate that there always exist the real WFs whose spread functional is less than any complex WFs.

## II. SOME REVIEWS

In this section, we review some physical concepts necessary for our work.

### A. Bloch wave function

Bloch's theorem states that in a periodic solid each electronic wave function can be written as the product of a cell-periodic part and a wavelike part,

$$\psi_{n,\mathbf{k}}(\mathbf{r}) = e^{i\mathbf{k}\cdot\mathbf{r}} \cdot u_{n,\mathbf{k}}(\mathbf{r}), \qquad (1)$$

where $\psi_{n,\mathbf{k}}$ is Bloch wave function and $n$, $\mathbf{k}$ stand for the band index and wave vector, respectively. The periodic part of the wave function can be expanded using a basis set consisting of a discrete set of plane waves whose wave vectors are reciprocal lattice vectors of the crystal,

$$u_{n,\mathbf{k}}(\mathbf{r}) = \sum_{\mathbf{G}} c_{n,\mathbf{k}}(\mathbf{G}) e^{i\mathbf{G}\cdot\mathbf{r}}, \qquad (2)$$

where the reciprocal lattice vectors $\mathbf{G}$ are defined by $\mathbf{G}\cdot\boldsymbol{l} = 2\pi m$ for all $\boldsymbol{l}$ where $\boldsymbol{l}$ is a lattice vector of the crystal and $m$ is an integer. Therefore, each electronic wave function can be written as a sum of plane waves,

$$\psi_{n,\mathbf{k}}(\mathbf{r}) = \sum_{\mathbf{G}} c_{n,\mathbf{k}}(\mathbf{G}) e^{i(\mathbf{k}+\mathbf{G})\cdot\mathbf{r}}. \qquad (3)$$

## B. Definition and measure of the spatial delocalization of Wannier functions

A Wannier function $w_n(\mathbf{r} - \mathbf{R})$, labeled by the lattice vector $\mathbf{R}$, is usually defined by a Fourier transform of the Bloch functions of the nth band:

$$w_n(\mathbf{r} - \mathbf{R}) = \frac{V}{(2\pi)^3} \int_{BZ} \psi_{n,\mathbf{k}}(\mathbf{r}) e^{-i\mathbf{k}\cdot\mathbf{R}} d^3\mathbf{k}, \quad (4)$$

where $V$ is the volume of the unit cell and the integration is performed over the first Brillouin zone. The transformation (4) assumes that the Bloch functions $\psi_{n,\mathbf{k}}(\mathbf{r})$ are periodic in reciprocal space, i. e. the "periodic gauge" where $\psi_{n,\mathbf{k}}(\mathbf{r}) = \psi_{n,\mathbf{k}+\mathbf{G}}(\mathbf{r})$. Since the overall phase of each wave function is arbitrary, any Bloch function is subjected to a "gauge transformation"

$$\psi_{n,\mathbf{k}}(\mathbf{r}) \to \tilde{\psi}_{n,\mathbf{k}}(\mathbf{r}) = e^{i\phi_n(\mathbf{k})} \psi_{n,\mathbf{k}}(\mathbf{r}), \quad (5)$$

which leaves unchanged all physically meaningful quantities. Besides this freedom in the choice of phases for the Bloch functions, there is a more comprehensive gauge freedom stemming from the fact that the many-body wave function is actually a Slater determinant: a unitary transformation between orbitals will not change the manifold, and will not change the total energy and the charge density of the system. In all generality, starting with a set of N Bloch functions, we can construct infinite sets of N WFs displaying different spatial characteristics:

$$w_n(\mathbf{r} - \mathbf{R}) = \frac{V}{(2\pi)^3} \int_{BZ} [\sum_m U_{mn}^{(\mathbf{k})} \psi_{m,\mathbf{k}}(\mathbf{r})] e^{-i\mathbf{k}\cdot\mathbf{r}} d^3\mathbf{k}. \quad (6)$$

The unitary matrix $U^{(\mathbf{k})}$ includes also the gauge freedom on phase factors. Finding the MLWFs means to search the unitary transformation $U^{(\mathbf{k})}$ that transform the Bloch wave functions in the WFs with the narrowest spatial distribution. A measure of the spatial delocalization of WFs is given by the spread functional $\Omega$, defined as the sum of the second moments of all the WFs in a reference cell:

$$\Omega = \sum_n \langle 0n|(\mathbf{r} - \bar{\mathbf{r}}_n)^2|0n\rangle = \sum_n [\langle r^2\rangle_n - \bar{\mathbf{r}}_n^2], \quad (7)$$

where the sum is over a selected group of bands, and

$$\langle r^2\rangle_n = \langle 0n|r^2|0n\rangle, \quad (8)$$

$$\bar{\mathbf{r}}_n = \langle 0n|\mathbf{r}|0n\rangle, \quad (9)$$

$$|0n\rangle = w_n(\mathbf{r}) = \frac{V}{(2\pi)^3} \int_{BZ} \psi_{n,\mathbf{k}}(\mathbf{r}) d^3\mathbf{k}. \quad (10)$$

## C. Localization procedure

Following the procedure proposed by Marzari and Vanderbilt, the expressions for $\langle r^2\rangle_n$ and $\bar{\mathbf{r}}_n$ are as follows:

$$\langle r^2\rangle_n = \frac{1}{N}\sum_{\mathbf{k},\mathbf{b}} \omega_\mathbf{b}\{1 - |M_{nn}^{(\mathbf{k},\mathbf{b})}|^2 + [argM_{nn}^{(\mathbf{k},\mathbf{b})}]^2\}, \quad (11)$$

$$\bar{\mathbf{r}}_n = -\frac{1}{N}\sum_{\mathbf{k},\mathbf{b}} \omega_\mathbf{b} \mathbf{b}\, argM_{nn}^{(\mathbf{k},\mathbf{b})}, \quad (12)$$

where

$$M_{mn}^{(\mathbf{k},\mathbf{b})} = \langle u_{m,\mathbf{k}}|u_{n,\mathbf{k}+\mathbf{b}}\rangle, \quad (13)$$

$\mathbf{b}$ is a vector connecting a $\mathbf{k}$ point to one of its neighbors, $N$ is the number of $\mathbf{k}$ points in the Brillouin zone and $\omega_\mathbf{b}$ are the weights of the $\mathbf{b}$ vectors satisfying the completeness condition

$$\sum_\mathbf{b} \omega_\mathbf{b} b_\alpha b_\beta = \delta_{\alpha\beta}. \quad (14)$$

They obtained the expression for the gradient of spread functional:

$$\mathbf{G}^{(\mathbf{k})} = 4 \sum_{\mathbf{k},\mathbf{b}} \omega_\mathbf{b} \left( \frac{R^{(\mathbf{k},\mathbf{b})} - R^{(\mathbf{k},\mathbf{b})\dagger}}{2} - \frac{T^{(\mathbf{k},\mathbf{b})} + T^{(\mathbf{k},\mathbf{b})\dagger}}{2i} \right), \quad (15)$$

where

$$R_{mn}^{(\mathbf{k},\mathbf{b})} = M_{mn}^{(\mathbf{k},\mathbf{b})} M_{nn}^{(\mathbf{k},\mathbf{b})*}, \quad (16)$$

$$T_{mn}^{(\mathbf{k},\mathbf{b})} = q_n^{(\mathbf{k},\mathbf{b})} \frac{M_{mn}^{(\mathbf{k},\mathbf{b})}}{M_{nn}^{(\mathbf{k},\mathbf{b})}} \quad (17)$$

with

$$q_n^{(\mathbf{k},\mathbf{b})} = argM_{nn}^{(\mathbf{k},\mathbf{b})} + \bar{\mathbf{r}}_n \cdot \mathbf{b}. \quad (18)$$

## D. Inversion-symmetry of density in momentum space

In general, the Hamiltonian $H = p^2/2m + V(\mathbf{r})$ that has the (real) potential $V(\mathbf{r})$ is real. From this, there exists a relationship between the Bloch wave functions. We'll review it here. The wave equation of a given system could be solved in the matrix form:

$$\sum_\mathbf{G} H_{\mathbf{G}'\mathbf{G}}^{(\mathbf{k})} c_{n,\mathbf{k}}(\mathbf{G}) = \varepsilon_{n,\mathbf{k}}\, c_{n,\mathbf{k}}(\mathbf{G}'), \quad (19)$$

where $\varepsilon_{n,\mathbf{k}}$ is the eigen-value for $n, \mathbf{k}$ and $H_{\mathbf{G}'\mathbf{G}}^{(\mathbf{k})}$ is the Hamiltonian matrix with

$$H_{\mathbf{G}'\mathbf{G}}^{(\mathbf{k})} = \int e^{-i(\mathbf{k}+\mathbf{G}')\cdot\mathbf{r}} \cdot H \cdot e^{i(\mathbf{k}+\mathbf{G})\cdot\mathbf{r}} d\mathbf{r}. \quad (20)$$

Assuming that the Hamiltonian is Hermitian and real, it is easy to check that

$$H_{-\mathbf{G}'-\mathbf{G}}^{(-\mathbf{k})} = H_{\mathbf{G}\mathbf{G}'}^{(\mathbf{k})}, \quad (21)$$

inserting $-\mathbf{k}, -\mathbf{G}, -\mathbf{G}'$ in place of $\mathbf{k}, \mathbf{G}, \mathbf{G}'$ into the equation (20). Inserting $-\mathbf{k}, -\mathbf{G}, -\mathbf{G}'$ in place of $\mathbf{k}, \mathbf{G}, \mathbf{G}'$ into the complex conjugate of the equation (19) and considering the equation (21), we obtain

$$H_{G'G}^{(k)} c_{n,-k}^*(-G) = \varepsilon_{n,-k} c_{n,-k}^*(-G), \qquad (22)$$

where $*$ means the complex conjugate. From the equation (19) and (22), we find that $c_{n,k}(G)$ equal to $c_{n,-k}^*(-G)$ for any $n, k, G$ within gauge accuracy:

$$c_{n,-k}^*(-G) = \lambda c_{n,k}(G) \text{ or } u_{n,-k}^*(r) = \lambda u_{n,k}(r), \qquad (23)$$

where $\lambda$ is an arbitrary complex whose absolute value is 1. From the equation (23) we see that the charge density has inversion-symmetry in momentum space even if the inversion symmetry is not present in the given system [29]:

$$\rho_k = \rho_{-k}. \qquad (24)$$

The equation (24) tells us that it is possible to use the equation

$$u_{n,-k}^*(r) = u_{n,k}(r) \text{ or } \psi_{n,-k}^*(r) = \psi_{n,k}(r) \qquad (25)$$

in the calculation of MLWFs because the density leaves unchanged.

### III. PRELIMINARY DISCUSSION

#### A. Description of global minima

Marzari and Vanderbilt point out that at a global minimum, all of the

$$|q_n^{(k,b)}| \ll \pi. \qquad (26)$$

The reason could be explained as follows: Considering the condition (14), we obtain the expression for the spread functional

$$\Omega = \frac{1}{N}\sum_{n,k,b} \omega_b \{1 - |M_{nn}^{(k,b)}|^2 + [q_n^{(k,b)}]^2\} \qquad (27)$$

from the equations (11) and (12). The equation (27) could be also directly obtained from the $\Omega_{I,OD}$ and $\Omega_D$ in the reference [6]. From the equation (27), it is evident that the spread functional arrives at a minimum when all of the $|q_n^{(k,b)}| = 0$.

#### B. Expression for spread functional by plane waves

Inserting equation (3) into equation (10), we get

$$w_n(r) = \sum_G w_{n,G}(r), \qquad (28)$$

where

$$w_{n,G}(r) = \frac{V}{(2\pi)^3} \int_{BZ} c_{n,k}(G) e^{i(k+G)\cdot r} d^3k. \qquad (29)$$

Inserting equation (28) into equation (8) and equation (9), we obtain

$$\langle r^2 \rangle_n = \frac{V^2}{(2\pi)^6} \sum_{GG'} I_{r^2}(G, G'), \qquad (30)$$

$$\bar{r}_n = \frac{V^2}{(2\pi)^6} \sum_{GG'} I_r(G, G'), \qquad (31)$$

where

$$I_{r^2}(G, G') = \iiint r^2 c_{n,k}(G) c_{n,k'}^*(G') e^{i(G-G'+k-k')\cdot r} d^3r d^3k d^3k', \qquad (32)$$

$$I_r(G, G') = \iiint r\, c_{n,k}(G) c_{n,k'}^*(G') e^{i(G-G'+k-k')\cdot r} d^3r d^3k d^3k'. \qquad (33)$$

In equations (30) and (31), we can write in the case of $G \neq G'$ as follows:

$$I_{r^2}(G, G') + I_{r^2}(G', G) = \iiint 2 r^2 |c_{n,k}(G)||c_{n,k'}(G')| \cos[\varphi_{n,k}(G) - \varphi_{n,k'}(G')] \cos[(G - G' + k - k')\cdot r] d^3r d^3k d^3k', \qquad (34)$$

$$I_r(G, G') + I_r(G', G) = -\iiint 2 r |c_{n,k}(G)||c_{n,k'}(G')| \sin[\varphi_{n,k}(G) - \varphi_{n,k'}(G')] \sin[(G - G' + k - k')\cdot r] d^3r d^3k d^3k'. \qquad (35)$$

And otherwise ($G = G'$) as follows:

$$I_{r^2}(G, G) = \iiint r^2 |c_{n,k}(G)||c_{n,k'}(G)| \cos[\varphi_{n,k}(G) - \varphi_{n,k'}(G)] \cos[(k - k')\cdot r] d^3r d^3k d^3k', \qquad (36)$$

$$I_r(G, G) = -\iiint r |c_{n,k}(G)||c_{n,k'}(G)| \sin[\varphi_{n,k}(G) - \varphi_{n,k'}(G)] \sin[(k - k')\cdot r] d^3r d^3k d^3k', \qquad (37)$$

where $|c_{n,k}(G)|$ and $\varphi_{n,k}(G)$ means the absolute value and phase of $c_{n,k}(G)$, respectively.

All the integrands in $\langle r^2 \rangle_n$ contain *cosine* with respect to $\varphi_{n,k}(G)$ for any $n, k, G$ and those in $\bar{r}_n$ include *sine*. Therefore, all the integrands in the spread functional contain *cosine*, noting that the product of *sine* is equal to the difference of *cosine*.

When inversion symmetry is present, the coefficients of plane waves of Bloch functions can be chosen to be real. Then, we can see that the spread functional of the corresponding WFs is stationary and $\bar{r}_n = 0$ from the equations (34)~(37) and all of the $q_n^{(k,b)} = 0$.

## IV. MAIN PROOF

Let $w_n^{(real)}$ and $w_n^{(cmplx)}$ be real WFs and complex WFs infinitesimally-deviated from real, respectively. Then, we write

$$w_n^{(cmplx)} = w_n^{(real)} + i\delta w_n, \tag{38}$$

where $\delta w_n$ is the infinitesimal function. Using the equation (38), we can estimate the spread functional :

$$\Omega^{(cmplx)} = \sum_n \langle w_n^{(cmplx)} | (\mathbf{r} - \bar{\mathbf{r}}_n^{(cmplx)})^2 | w_n^{(cmplx)} \rangle$$

$$= \sum_n \langle w_n^{(real)} | (\mathbf{r} - \bar{\mathbf{r}}_n^{(cmplx)})^2 | w_n^{(real)} \rangle$$
$$+ \sum_n \langle \delta w_n, | (\mathbf{r} - \bar{\mathbf{r}}_n^{(cmplx)})^2 | \delta w_n, \rangle$$

$$> \sum_n \langle w_n^{(real)} | (\mathbf{r} - \bar{\mathbf{r}}_n^{(cmplx)})^2 | w_n^{(real)} \rangle$$

$$= \sum_n \langle w_n^{(real)} | (\mathbf{r} - \bar{\mathbf{r}}_n^{(real)} - \delta \bar{\mathbf{r}}_n)^2 | w_n^{(real)} \rangle$$

$$= \sum_n \langle w_n^{(real)} | (\mathbf{r} - \bar{\mathbf{r}}_n^{(real)})^2 | w_n^{(real)} \rangle$$
$$+ \sum_n (\delta \bar{\mathbf{r}}_n)^2 \langle w_n^{(real)} | w_n^{(real)} \rangle$$

$$> \sum_n \langle w_n^{(real)} | (\mathbf{r} - \bar{\mathbf{r}}_n^{(real)})^2 | w_n^{(real)} \rangle = \Omega^{(real)}, \tag{39}$$

where $\Omega^{(cmplx)}$ and $\Omega^{(real)}$ are spread functional of complex and real WFs, respectively, $\bar{\mathbf{r}}_n^{(cmplx)}$ and $\bar{\mathbf{r}}_n^{(real)}$ are Wannier centers of $w_n^{(cmplx)}$ and $w_n^{(real)}$, and $\delta \bar{\mathbf{r}}_n = \bar{\mathbf{r}}_n^{(cmplx)} - \bar{\mathbf{r}}_n^{(real)}$. It is used that

$$\langle w_n^{(real)} | (\mathbf{r} - \bar{\mathbf{r}}_n^{(real)}) | w_n^{(real)} \rangle = 0. \tag{40}$$

From the equation (39), we see that they'll trap in the real if there exist the real WFs around the complex WFs during the minimization procedure. Of course, if not exist, they will arrive at the complex.

In the next part, we'll state that the minimized complex WFs are false local minima of the spread functional. To do this, we demonstrate that there always exist the real WFs whose spread functional is less than any complex.

The WFs projected from the Bloch wave functions that the equation (25) is satisfied are real:

$$w_n(\mathbf{r}) =$$
$$\frac{V}{(2\pi)^3} [\int_{(BZ/2)^+} e^{i\mathbf{k}\cdot\mathbf{r}} \cdot u_{n,\mathbf{k}}(\mathbf{r}) \, d^3\mathbf{k}$$
$$+ \int_{(BZ/2)^-} e^{i\mathbf{k}\cdot\mathbf{r}} \cdot u_{n,\mathbf{k}}(\mathbf{r}) \, d^3\mathbf{k}]$$

$$= \frac{V}{(2\pi)^3} [\int_{(BZ/2)^+} e^{i\mathbf{k}\cdot\mathbf{r}} \cdot u_{n,\mathbf{k}}(\mathbf{r}) \, d^3\mathbf{k}$$
$$+ \int_{(BZ/2)^+} e^{-i\mathbf{k}\cdot\mathbf{r}} \cdot u_{n,-\mathbf{k}}(\mathbf{r}) \, d^3\mathbf{k}]$$

$$= \frac{V}{(2\pi)^3} \int_{(BZ/2)^+} [e^{i\mathbf{k}\cdot\mathbf{r}} \cdot u_{n,\mathbf{k}}(\mathbf{r}) + e^{-i\mathbf{k}\cdot\mathbf{r}} \cdot u_{n,\mathbf{k}}^*(\mathbf{r})] d^3\mathbf{k} \rightarrow \text{real} \tag{41}$$

or

$$w_n(\mathbf{r}) =$$
$$\frac{V}{(2\pi)^3} [\int_{(BZ/2)} e^{-i\mathbf{k}\cdot\mathbf{r}} \cdot u_{n,-\mathbf{k}}(\mathbf{r}) \, d^3\mathbf{k}$$
$$+ \int_{(BZ/2)^-} e^{i\mathbf{k}\cdot\mathbf{r}} \cdot u_{n,\mathbf{k}}(\mathbf{r}) \, d^3\mathbf{k}]$$

$$= \frac{V}{(2\pi)^3} \int_{(BZ/2)^-} [e^{-i\mathbf{k}\cdot\mathbf{r}} \cdot u_{n,\mathbf{k}}^*(\mathbf{r}) + e^{i\mathbf{k}\cdot\mathbf{r}} \cdot u_{n,\mathbf{k}}(\mathbf{r})] d^3\mathbf{k}$$

$$= \frac{V}{(2\pi)^3} \int_{(BZ/2)^+} [e^{i\mathbf{k}\cdot\mathbf{r}} \cdot u_{n,\mathbf{k}}(\mathbf{r}) + e^{-i\mathbf{k}\cdot\mathbf{r}} \cdot u_{n,\mathbf{k}}^*(\mathbf{r})] d^3\mathbf{k} \rightarrow \text{real} \tag{42}$$

where $(BZ/2)^+$ and $(BZ/2)^-$ are haves of Brillouin zone with opposite signs of $\mathbf{k}$, respectively. From the equation (41) or (42), we see that the real WFs could be constructed with the Bloch functions calculated in half of Brillouin zone

Let's think a deviation from the equation (25). The simultaneous infinitesimal transformation, $\varepsilon$ of $\varphi_{n,\mathbf{k}}(\mathbf{G})$ and $\varphi_{n,-\mathbf{k}}(-\mathbf{G})$ makes the difference between the absolute values of $\varphi_{n,\mathbf{k}}(\mathbf{G})$ and $\varphi_{n,-\mathbf{k}}(-\mathbf{G})$ (phase of $\lambda$ in equation (23)) to be $2\varepsilon$. Considering the terms with opposite sign of $\mathbf{k}$ and $\mathbf{G}$ in the equations (34~37), one possible stationary solution of $\Omega$ by the above transformation is that all the coefficients satisfy the equation (25). With this meaning, we can consider that such real WFs are closer to the stationary solution.

If the WFs are complex, all of $u_{n,\mathbf{k}}(\mathbf{r})$ do not satisfy the equation (25). We denote the Bloch functions calculated in $(BZ/2)^+$ and $(BZ/2)^-$, $u_{n,\mathbf{k}}(\mathbf{r})$ and $v_{n,\mathbf{k}}(\mathbf{r})$, respectively. Then, from the $u_{n,\mathbf{k}}(\mathbf{r})$ and $v_{n,\mathbf{k}}(\mathbf{r})$, we can obtain two real WFs set, $\text{Re}w_n^{(1)}(\mathbf{r})$ and $\text{Re}w_n^{(2)}(\mathbf{r})$, where

$$\text{Re}w_n^{(1)} =$$
$$\frac{V}{(2\pi)^3} \int_{(BZ/2)^+} [e^{i\mathbf{k}\cdot\mathbf{r}} \cdot u_{n,\mathbf{k}}(\mathbf{r}) + e^{-i\mathbf{k}\cdot\mathbf{r}} \cdot u_{n,\mathbf{k}}^*(\mathbf{r})] d^3\mathbf{k}$$

$$= \frac{V}{(2\pi)^3} \int_{(BZ/2)^+} e^{i\mathbf{k}\cdot\mathbf{r}} \cdot u_{n,\mathbf{k}}(\mathbf{r}) \, d^3\mathbf{k}$$
$$+ \frac{V}{(2\pi)^3} \int_{(BZ/2)^-} e^{i\mathbf{k}\cdot\mathbf{r}} \cdot u_{n,-\mathbf{k}}^*(\mathbf{r}) \, d^3\mathbf{k}$$

$$= \frac{V}{(2\pi)^3} \int_{BZ} e^{i\mathbf{k}\cdot\mathbf{r}} \cdot u_{n,\mathbf{k}}(\mathbf{r}) \, d^3\mathbf{k} \tag{43}$$

and

$$\text{Re}w_n^{(2)} =$$
$$\frac{V}{(2\pi)^3} \int_{(BZ/2)^-} [e^{i\mathbf{k}\cdot\mathbf{r}} \cdot v_{n,\mathbf{k}}(\mathbf{r}) + e^{-i\mathbf{k}\cdot\mathbf{r}} \cdot v_{n,\mathbf{k}}^*(\mathbf{r})] d^3\mathbf{k}$$

$$= \frac{V}{(2\pi)^3} \int_{(BZ/2)^-} e^{i\mathbf{k}\cdot\mathbf{r}} \cdot v_{n,\mathbf{k}}(\mathbf{r}) \, d^3\mathbf{k}$$

$$+ \frac{V}{(2\pi)^3} \int_{(BZ/2)^+} e^{i\mathbf{k}\cdot\mathbf{r}} \cdot v_{n,-\mathbf{k}}^*(\mathbf{r})\, d^3\mathbf{k}$$

$$= \frac{V}{(2\pi)^3} \int_{BZ} e^{i\mathbf{k}\cdot\mathbf{r}} \cdot v_{n,\mathbf{k}}(\mathbf{r})\, d^3\mathbf{k} \quad (44)$$

with $u_{n,\mathbf{k}}(\mathbf{r}) = u_{n,-\mathbf{k}}^*(\mathbf{r})$ and $v_{n,\mathbf{k}}(\mathbf{r}) = v_{n,-\mathbf{k}}^*(\mathbf{r})$, respectively.

The real part of any complex WFs, $w_n(\mathbf{r})$ are expressed as the algebraic average of two real WFs:

$$(w_n(\mathbf{r}) + w_n^*(\mathbf{r}))/2 =$$

$$= \frac{V}{2(2\pi)^3} \int_{(BZ/2)^+} e^{i\mathbf{k}\cdot\mathbf{r}} \cdot u_{n,\mathbf{k}}(\mathbf{r})\, d^3\mathbf{k}$$
$$+ \frac{V}{2(2\pi)^3} \int_{(BZ/2)^-} e^{i\mathbf{k}\cdot\mathbf{r}} \cdot v_{n,\mathbf{k}}(\mathbf{r})\, d^3\mathbf{k}$$
$$+ \frac{V}{2(2\pi)^3} \int_{(BZ/2)^+} e^{-i\mathbf{k}\cdot\mathbf{r}} \cdot u_{n,\mathbf{k}}^*(\mathbf{r})\, d^3\mathbf{k}$$
$$+ \frac{V}{2(2\pi)^3} \int_{(BZ/2)^-} e^{-i\mathbf{k}\cdot\mathbf{r}} \cdot v_{n,\mathbf{k}}^*(\mathbf{r})\, d^3\mathbf{k}$$

$$= (\mathrm{Re}\, w_n^{(1)}(\mathbf{r}) + \mathrm{Re}\, w_n^{(2)}(\mathbf{r}))/2 \quad (45)$$

We can also obtain the imaginary part of $w_n(\mathbf{r})$:

$$i(w_n^*(\mathbf{r}) - w_n(\mathbf{r}))/2$$

$$= \frac{Vi}{2(2\pi)^3} \int_{(BZ/2)^+} e^{-i\mathbf{k}\cdot\mathbf{r}} \cdot u_{n,\mathbf{k}}^*(\mathbf{r})\, d^3\mathbf{k}$$
$$+ \frac{Vi}{2(2\pi)^3} \int_{(BZ/2)^-} e^{-i\mathbf{k}\cdot\mathbf{r}} \cdot v_{n,\mathbf{k}}^*(\mathbf{r})\, d^3\mathbf{k}$$
$$- \frac{Vi}{2(2\pi)^3} \int_{(BZ/2)^+} e^{i\mathbf{k}\cdot\mathbf{r}} \cdot u_{n,\mathbf{k}}(\mathbf{r})\, d^3\mathbf{k}$$
$$- \frac{Vi}{2(2\pi)^3} \int_{(BZ/2)^-} e^{i\mathbf{k}\cdot\mathbf{r}} \cdot v_{n,\mathbf{k}}(\mathbf{r})\, d^3\mathbf{k}$$

$$= \frac{Vi}{2(2\pi)^3} \int_{(BZ/2)^+} e^{-i\mathbf{k}\cdot\mathbf{r}} \cdot u_{n,\mathbf{k}}^*(\mathbf{r})\, d^3\mathbf{k}$$
$$+ \frac{Vi}{2(2\pi)^3} \int_{(BZ/2)^+} e^{i\mathbf{k}\cdot\mathbf{r}} \cdot v_{n,\mathbf{k}}(\mathbf{r})\, d^3\mathbf{k}$$
$$- \frac{Vi}{2(2\pi)^3} \int_{(BZ/2)^+} e^{i\mathbf{k}\cdot\mathbf{r}} \cdot u_{n,\mathbf{k}}(\mathbf{r})\, d^3\mathbf{k}$$
$$- \frac{Vi}{2(2\pi)^3} \int_{(BZ/2)^+} e^{-i\mathbf{k}\cdot\mathbf{r}} \cdot v_{n,\mathbf{k}}^*(\mathbf{r})\, d^3\mathbf{k}$$

$$= i(wd_n^*(\mathbf{r}) - wd_n(\mathbf{r}))/2 = \mathrm{Im}\, wd_n(\mathbf{r}), \quad (46)$$

where

$$wd_n(\mathbf{r}) = \frac{V}{(2\pi)^3} \int_{(BZ/2)^+} e^{i\mathbf{k}\cdot\mathbf{r}} (u_{n,\mathbf{k}} - v_{n,\mathbf{k}})\, d^3\mathbf{k}. \quad (47)$$

We can express the difference of two real WFs by the equation (47):

$$(\mathrm{Re}\, w_n^{(1)}(\mathbf{r}) - \mathrm{Re}\, w_n^{(2)}(\mathbf{r}))/2$$

$$= \frac{V}{2(2\pi)^3} \int_{(BZ/2)^+} e^{i\mathbf{k}\cdot\mathbf{r}} \cdot u_{n,\mathbf{k}}(\mathbf{r})\, d^3\mathbf{k}$$
$$+ \frac{V}{2(2\pi)^3} \int_{(BZ/2)^+} e^{-i\mathbf{k}\cdot\mathbf{r}} \cdot u_{n,\mathbf{k}}^*(\mathbf{r})\, d^3\mathbf{k}$$
$$- \frac{V}{2(2\pi)^3} \int_{(BZ/2)^+} e^{i\mathbf{k}\cdot\mathbf{r}} \cdot v_{n,\mathbf{k}}(\mathbf{r})\, d^3\mathbf{k}$$

$$- \frac{V}{2(2\pi)^3} \int_{(BZ/2)^+} e^{-i\mathbf{k}\cdot\mathbf{r}} \cdot v_{n,\mathbf{k}}^*(\mathbf{r})\, d^3\mathbf{k}$$

$$= (wd_n(\mathbf{r}) + wd_n^*(\mathbf{r}))/2 = \mathrm{Re}\, wd_n(\mathbf{r}) \quad (48)$$

Considering the equations (45) and (46), we write

$$w_n(\mathbf{r}) = (\mathrm{Re}\, w_n^{(1)}(\mathbf{r}) + \mathrm{Re}\, w_n^{(2)}(\mathbf{r}))/2 + i\, \mathrm{Im}\, wd_n(\mathbf{r})$$
$$= \mathrm{Re}\, w_n(\mathbf{r}) + i\, \mathrm{Im}\, wd_n(\mathbf{r}), \quad (49)$$

where $\mathrm{Re}\, w_n = (\mathrm{Re}\, w_n^{(1)} + \mathrm{Re}\, w_n^{(2)})/2$. Let's assume that

$$\left\langle \mathrm{Re}\, w_n^{(1)} \middle| (\mathbf{r} - \bar{\mathbf{r}}_{n,c})^2 \middle| \mathrm{Re}\, w_n^{(1)} \right\rangle \leq$$
$$\left\langle \mathrm{Re}\, w_n^{(2)} \middle| (\mathbf{r} - \bar{\mathbf{r}}_{n,c})^2 \middle| \mathrm{Re}\, w_n^{(2)} \right\rangle, \quad (50)$$

where $\bar{\mathbf{r}}_{n,c} = \langle \mathrm{Re}\, w_n | \mathbf{r} | \mathrm{Re}\, w_n \rangle + \langle \mathrm{Im}\, wd_n | \mathbf{r} | \mathrm{Im}\, wd_n \rangle$. We get from the inequality (50)

$$2 * \left\langle \mathrm{Re}\, w_n^{(1)} \middle| (\mathbf{r} - \bar{\mathbf{r}}_{n,c})^2 \middle| \mathrm{Re}\, w_n^{(2)} - \mathrm{Re}\, w_n^{(1)} \right\rangle +$$
$$\left\langle \mathrm{Re}\, w_n^{(2)} - \mathrm{Re}\, w_n^{(1)} \middle| (\mathbf{r} - \bar{\mathbf{r}}_{n,c})^2 \middle| \mathrm{Re}\, w_n^{(2)} - \mathrm{Re}\, w_n^{(1)} \right\rangle \geq 0. \quad (51)$$

Using the inequality (51), we can estimate the spread functional of the complex WFs for any band, $\mathbf{\Omega}(w_n)$:

$$\mathbf{\Omega}(w_n) = \left\langle \mathrm{Re}\, w_n \middle| (\mathbf{r} - \bar{\mathbf{r}}_{n,c})^2 \middle| \mathrm{Re}\, w_n \right\rangle$$
$$+ \left\langle \mathrm{Im}\, wd_n \middle| (\mathbf{r} - \bar{\mathbf{r}}_{n,c})^2 \middle| \mathrm{Im}\, wd_n \right\rangle$$

$$= \left\langle \mathrm{Re}\, w_n^{(1)} \middle| (\mathbf{r} - \bar{\mathbf{r}}_{n,c})^2 \middle| \mathrm{Re}\, w_n^{(1)} \right\rangle$$
$$+ \left\langle \mathrm{Re}\, w_n^{(1)} \middle| (\mathbf{r} - \bar{\mathbf{r}}_{n,c})^2 \middle| \mathrm{Re}\, w_n^{(2)} - \mathrm{Re}\, w_n^{(1)} \right\rangle$$
$$+ \frac{1}{4} \left\langle \mathrm{Re}\, w_n^{(2)} - \mathrm{Re}\, w_n^{(1)} \middle| (\mathbf{r} - \bar{\mathbf{r}}_{n,c})^2 \middle| \mathrm{Re}\, w_n^{(2)} - \mathrm{Re}\, w_n^{(1)} \right\rangle$$
$$+ \left\langle \mathrm{Im}\, wd_n \middle| (\mathbf{r} - \bar{\mathbf{r}}_{n,c})^2 \middle| \mathrm{Im}\, wd_n \right\rangle \geq$$

$$\left\langle \mathrm{Re}\, w_n^{(1)} \middle| (\mathbf{r} - \bar{\mathbf{r}}_{n,r})^2 \middle| \mathrm{Re}\, w_n^{(1)} \right\rangle + (\Delta \bar{\mathbf{r}}_{n,r})^2 \left\langle \mathrm{Re}\, w_n^{(1)} \middle| \mathrm{Re}\, w_n^{(1)} \right\rangle$$
$$- \left\langle (\mathrm{Re}\, w_n^{(2)} - \mathrm{Re}\, w_n^{(1)})/2 \middle| (\mathbf{r} - \bar{\mathbf{r}}_{n,c})^2 \middle| (\mathrm{Re}\, w_n^{(2)} - \mathrm{Re}\, w_n^{(1)})/2 \right\rangle$$
$$+ \left\langle \mathrm{Im}\, w_n \middle| (\mathbf{r} - \bar{\mathbf{r}}_{n,c})^2 \middle| \mathrm{Im}\, w_n \right\rangle$$

$$\geq \left\langle \mathrm{Re}\, w_n^{(1)} \middle| (\mathbf{r} - \bar{\mathbf{r}}_{n,r})^2 \middle| \mathrm{Re}\, w_n^{(1)} \right\rangle$$
$$- \left\langle \mathrm{Re}\, wd_n \middle| (\mathbf{r} - \bar{\mathbf{r}}_{n,c})^2 \middle| \mathrm{Re}\, wd_n \right\rangle$$
$$+ \left\langle \mathrm{Im}\, wd_n \middle| (\mathbf{r} - \bar{\mathbf{r}}_{n,c})^2 \middle| \mathrm{Im}\, wd_n \right\rangle, \quad (52)$$

where $\bar{\mathbf{r}}_{n,r} = \left\langle \mathrm{Re}\, w_n^{(1)} \middle| \mathbf{r} \middle| \mathrm{Re}\, w_n^{(1)} \right\rangle$ and $\Delta \bar{\mathbf{r}}_{n,r} = \bar{\mathbf{r}}_{n,c} - \bar{\mathbf{r}}_{n,r}$. In the case of

$$\langle \operatorname{Re} wd_n | (\mathbf{r} - \bar{\mathbf{r}}_{n,c})^2 | \operatorname{Re} wd_n \rangle$$
$$\leq \langle \operatorname{Im} wd_n | (\mathbf{r} - \bar{\mathbf{r}}_{n,c})^2 | \operatorname{Im} wd_n \rangle, \tag{53}$$

from the inequality (52), we get

$$\Omega(w_n) \geq \Omega(\operatorname{Re} w_n^{(1)}), \tag{54}$$

where $\Omega(\operatorname{Re} w_n^{(1)}) = \langle \operatorname{Re} w_n^{(1)} | (\mathbf{r} - \bar{\mathbf{r}}_{n,r})^2 | \operatorname{Re} w_n^{(1)} \rangle$.

To investigate the case of

$$\langle \operatorname{Re} wd_n | (\mathbf{r} - \bar{\mathbf{r}}_{n,c})^2 | \operatorname{Re} wd_n \rangle$$
$$\geq \langle \operatorname{Im} wd_n | (\mathbf{r} - \bar{\mathbf{r}}_{n,c})^2 | \operatorname{Im} wd_n \rangle, \tag{55}$$

we write the inequality (52) for $w_n'(\mathbf{r}) = iw_n(\mathbf{r})$ with $u'_{n,\mathbf{k}} = iu_{n,\mathbf{k}}$ and $v'_{n,\mathbf{k}} = iv_{n,\mathbf{k}}$,

$$\Omega(w_n) = \Omega(w_n')$$
$$\geq \langle \operatorname{Re} w_n'^{(1)} | (\mathbf{r} - \bar{\mathbf{r}}'_{n,r})^2 | \operatorname{Re} w_n'^{(1)} \rangle$$
$$- \langle \operatorname{Re} wd_n' | (\mathbf{r} - \bar{\mathbf{r}}'_{n,c})^2 | \operatorname{Re} wd_n' \rangle$$
$$+ \langle \operatorname{Im} wd_n' | (\mathbf{r} - \bar{\mathbf{r}}'_{n,c})^2 | \operatorname{Im} wd_n' \rangle, \tag{56}$$

where

$$w_n'^{(1)} = \frac{V}{(2\pi)^3} \int_{(BZ/2)^+} [e^{i\mathbf{k}\cdot\mathbf{r}} \cdot u'_{n,\mathbf{k}}(\mathbf{r}) + e^{-i\mathbf{k}\cdot\mathbf{r}} \cdot u'^*_{n,\mathbf{k}}(\mathbf{r})] d^3\mathbf{k}$$
$$\frac{Vi}{(2\pi)^3} \int_{(BZ/2)^+} [e^{i\mathbf{k}\cdot\mathbf{r}} u_{n,\mathbf{k}}(\mathbf{r}) - e^{-i\mathbf{k}\cdot\mathbf{r}} \cdot u^*_{n,\mathbf{k}}(\mathbf{r})] d^3\mathbf{k}, \tag{57}$$

$$wd_n' = \frac{V}{(2\pi)^3} \int_{(BZ/2)^+} e^{i\mathbf{k}\cdot\mathbf{r}} (u'_{n,\mathbf{k}} - v'_{n,\mathbf{k}}) d^3\mathbf{k}$$
$$= \frac{Vi}{(2\pi)^3} \int_{(BZ/2)^+} e^{i\mathbf{k}\cdot\mathbf{r}} (u_{n,\mathbf{k}} - v_{n,\mathbf{k}}) d^3\mathbf{k} = iwd_n, \tag{58}$$

$$\bar{\mathbf{r}}'_{n,c} = \langle w_n' | \mathbf{r} | w_n' \rangle = \langle w_n | \mathbf{r} | w_n \rangle = \bar{\mathbf{r}}_{n,c}, \tag{59}$$

and

$$\bar{\mathbf{r}}'_{n,r} = \langle \operatorname{Re} w_n'^{(1)} | \mathbf{r} | \operatorname{Re} w_n'^{(1)} \rangle. \tag{60}$$

From the equation (58), we get

$$\operatorname{Re} wd_n' = -\operatorname{Im} wd_n, \quad \operatorname{Im} wd_n' = \operatorname{Re} wd_n. \tag{61}$$

Considering the equations (55), (59) and (61), we can rewrite the inequality (56)

$$\Omega(w_n) \geq$$
$$\langle \operatorname{Re} w_n'^{(1)} | (\mathbf{r} - \bar{\mathbf{r}}'_{n,r})^2 | \operatorname{Re} w_n'^{(1)} \rangle$$
$$- \langle \operatorname{Im} wd_n | (\mathbf{r} - \bar{\mathbf{r}}_{n,c})^2 | \operatorname{Im} wd_n \rangle$$
$$+ \langle \operatorname{Re} wd_n | (\mathbf{r} - \bar{\mathbf{r}}_{n,c})^2 | \operatorname{Re} wd_n \rangle$$
$$\geq \Omega(\operatorname{Re} w_n'^{(1)}), \tag{62}$$

where $\Omega(\operatorname{Re} w_n'^{(1)}) = \langle \operatorname{Re} w_n'^{(1)} | (\mathbf{r} - \bar{\mathbf{r}}'_{n,r})^2 | \operatorname{Re} w_n'^{(1)} \rangle$.

From the equations (50), (54) and (62), we can see that the spread functional of complex WFs is more than one of the following 4 real WFs that the corresponding Bloch functions are *eigen*-solutions of the given system

$$\frac{V}{(2\pi)^3} \int_{(BZ/2)^+} [e^{i\mathbf{k}\cdot\mathbf{r}} \cdot u_{n,\mathbf{k}}(\mathbf{r}) + e^{-i\mathbf{k}\cdot\mathbf{r}} \cdot u^*_{n,\mathbf{k}}(\mathbf{r})] d^3\mathbf{k},$$
$$\frac{Vi}{(2\pi)^3} \int_{(BZ/2)^+} [e^{i\mathbf{k}\cdot\mathbf{r}} \cdot u_{n,\mathbf{k}}(\mathbf{r}) - e^{-i\mathbf{k}\cdot\mathbf{r}} \cdot u^*_{n,\mathbf{k}}(\mathbf{r})] d^3\mathbf{k},$$
$$\frac{V}{(2\pi)^3} \int_{(BZ/2)^-} [e^{i\mathbf{k}\cdot\mathbf{r}} \cdot v_{n,\mathbf{k}}(\mathbf{r}) + e^{-i\mathbf{k}\cdot\mathbf{r}} \cdot v^*_{n,\mathbf{k}}(\mathbf{r})] d^3\mathbf{k},$$
$$\frac{Vi}{(2\pi)^3} \int_{(BZ/2)^-} [e^{i\mathbf{k}\cdot\mathbf{r}} \cdot v_{n,\mathbf{k}}(\mathbf{r}) - e^{-i\mathbf{k}\cdot\mathbf{r}} \cdot v^*_{n,\mathbf{k}}(\mathbf{r})] d^3\mathbf{k}.$$

In other words, we see that there always exist the real WFs whose spread functional is less than any complex WFs. So, we conclude that MLWFs are real.

During the minimization procedure by Marzari and Vanderbilt [6], the equation (25) continues to be satisfied. So, in this case, it is sufficient that the gradients of spread functional are only calculated in half of Brillouin zone. From the equation (54) or (62), we can see that more localized WFs would be obtained by such a calculation.

Imposing the initial condition (25) does not eliminate false local minima. To arrive at a global minimum, both the initial condition (25) and (26) should be satisfied.

## V. SUMMARY

In this paper, we proved the conjecture that the MLWFs are real. The minimized complex WFs are false local minima of the spread functional because there always exist the real WFs whose spread functional is less than any complex WFs. In detail, the real parts of any complex WFs are represented by the algebraic average of two real WFs and the spread functional of one of two real WFs is less than one of the complex WFs. So, the MLWFs should be real. Because of time reversal symmetry of Hamiltonian, we can start with initial real WFs. Then, the gradients of the spread functional are only calculated in half of Brillouin zone and more localized WFs could be obtained.

The fact that the MLWFs are real is similar to mode-locked laser pulse in laser physics with the meaning that the phases are consistent and well localized.